\documentclass{PoS}
 
\title{A Global Study of X-ray Binaries}

\ShortTitle{Global Study of X-ray Binaries}

\author{\speaker{Robert Dunn}\\
        University of Southampton\thanks{Now at Excellence Cluster
        Universe, Technische Universit\"at M\"unchen}\\
        E-mail: \email{robert.dunn@universe-cluster.de}}

\author{Rob Fender\\
        University of Southampton\\
        Email: \email{r.fender@soton.ac.uk}}
\author{Elmar K\"ording\\
        University of Southampton\\
        E-mail: \email{elmar@phys.soton.ac.uk}}
\author{Clement Cabanac\\
        University of Southampton\\
        E-mail: \email{c.cabanac@astro.soton.ac.uk}}
\author{Tomaso Belloni\\
        INAF-Osservatorio Astronomico di Brera\\
        E-mail: \email{tomaso.belloni@brera.inaf.it}}

\abstract{We present preliminary results on a global study of X-ray
  binaries using 14 Ms of data from the Rossi X-ray Timing Explorer {\it RXTE} satellite.  Our initial
  study on {\it GX~339-4} is recapped as an introduction to the methods
  used.  We use a consistent analysis scheme for all objects, with
  three different spectral models to fit the powerlaw and disc
  components.  We also take into account the possibility of a line
  being present in the data.  The resulting almost 4000 observations
  allow the tracking of the spectral properties of the binaries as
  they evolve through an outburst.  Our investigations concentrate on
  the disc and line properties of the binaries when in outburst.  We
  also show the Disc-Fraction Luminosity diagram for the population of
  X-ray binaries studied which will enable us to further links with AGN.}

\FullConference{VII Microquasar Workshop: Microquasars and Beyond\\
		 September 1-5 2008\\
		 Foca, Izmir, Turkey}

\begin{document}

\section{Aim}

\noindent The aim of this study is to investigate the properties of the
population of X-ray binaries, focusing on their outbursts.  We start by applying a single data
reduction scheme to one black hole X-ray binary, and then extend this
to other black hole X-ray binaries whose outbursts have been observed
by {\it RXTE}.  We then plan to use the results of this study to
further the links of X-ray binaries with AGN.

\section{Data Reduction} \label{sec:intro}

\noindent The {\it RXTE} telescope has been observing the X-ray sky for the past
11 years, resulting in a long baseline to investigate the outbursts of
X-ray binaries.  We used all the public data on X-ray binaries in the
{\it RXTE} Archive 
available on the 29 July 2008, corresponding to 4085 observations or
15.47Ms from 19 X-ray binaries..  The data were re-reduced to
standardise the data-products for 
all observations rather than using the standard data-products.  The
data reduction and spectral analysis was scripted for consistency.

We use data from both {\it HEXTE} as well as {\it PCA}.  The high
energy sensitivity from {\it HEXTE} allows the high-energy power-law slope
to be more accurately constrained, which assists greatly when
investigating the disc, most of whose emission falls below the
sensitivity bounds of the {\it PCA}.  Of the five {\it PCA}
detectors, however only one ({\it PCU-2}) has been on for the entire
length of the 
mission, and so for consistency we use that.  We use data from both
{\it HEXTE} Clusters where reliable backgrounds could be created.

To allow for reliable, accurate and relatively quick spectral fitting
we do not fit spectra which have fewer than 1000 background subtracted
{\it PCA} counts.   The excluded observations occur throughout the
light-curves of the objects, with a concentration in low-flux
periods.  As our analysis concentrates on the outbursts of the
binaries, our results should not be biased by the exclusion of this
low counts data.  We also require that there are more than 2000
background subtracted counts from both {\it HEXTE} clusters combined
to allow the high energy powerlaw to be fitted accurately.  This in
turn means that the disc is more easily fitted.  The well calibrated energy range of
the {\it PCA} detectors extends to around $3$~keV at its lower bound.
The temperatures of the X-ray binary discs are around $1$~keV and so
being able to clearly fit the powerlaw component means that the disc
is more easily fit.

\section{Spectral Fitting}

\noindent We fit the spectra in {\scshape xspec}, using {\it PCA} absolute
channel number 7\footnote{Channel 7 varies from $2.27$ to $3.28$~kev
  over the lifetime of the mission.}  to an upper energy of
$25$~keV and {\it HEXTE} $25-250$~keV.  We use a channel rather than
an energetic cut-off for the {\it PCA} as this allows us to have a
well calibrated spectrum to the lowest possible energies; the energies
corresponding to channel number changed throughout the mission and all
channels greater than 6 are well calibrated.  The galactic absorption was
fixed to values taken from more detailed studies of the
binaries\footnote{The limitations on the low energy spectra from the
  {\it PCA} do not allow us to accurately fit the absorption as a free
parameter in the spectrum.}.  To
investigate the state of the X-ray binaries we fit three types of model:
{\scshape powerlaw}, {\scshape broken powerlaw} and {\scshape disc
  +powerlaw}\footnote{We investigated fitting comptonisation
  models.  However, only for the spectra with the largest numbers of
  counts was it possible to obtain accurate fits where no parameters
  required fixing.  As we wanted to have an automated routine, and
  many spectra had orders of magnitude fewer counts, we decided not to
pursue these models any further.}.   A version of each model also
having a Gaussian line at fixed energy of 
$6.4$~keV was fitted.  In fitting spectra of {\it
  GX~339-4}, when a broken powerlaw was selecting as the best fitting
model (126/700), 87 have $\Gamma_1>\Gamma_2$ and a break energy of
around $10$~keV and 39 have
$\Gamma_1<\Gamma_2$ with a break energy of around $40$~keV.  Most of
those with $\Gamma_1<\Gamma_2$ occur on the rise before the 2002/03 Outburst.

The best fitting model was selected on $\chi^2$ terms.  To test for
the significance of the line we used both an F-test and also made sure
that the line components normalisation was at least three times its
error.  We excluded all low flux observations $(<1\times
10^{11}$~erg/s$)$ as well as ones where the flux was not well
determined.  This resulted in a final list of 3680 observations
corresponding to 14.05Ms.

\section{GX~339-4}

\noindent In this Section we recap some of the results presented in
\cite{Dunn08} on the outbursts of {\it GX~339-4}.  The extra data
available in the archive subsequent to the initial study has been
included here.  

\subsection{Outbursts}

\begin{figure}
\includegraphics[width=7.5cm]{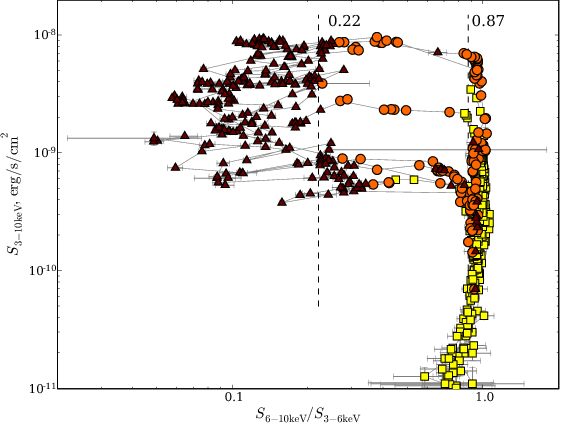}
\includegraphics[width=7.5cm]{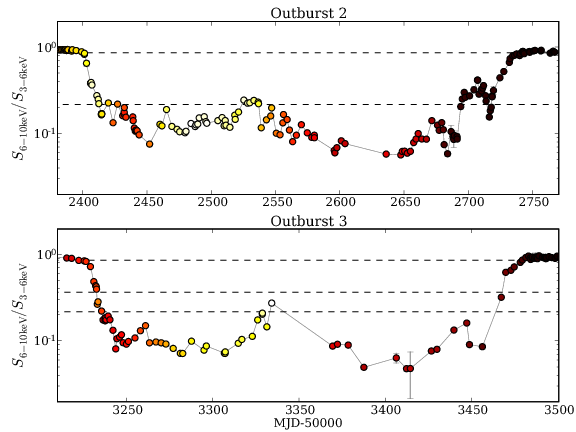}
\caption{{\scshape left} The HID for the outbursts from
{\it GX339-4}.  The different symbols correspond to the type of model
which resulted in the best fit.  Squares are the powerlaw, circles are the
broken powerlaw and triangles are the powerlaw+disc models.  The
dashed lines at 0.3 and 0.8 show the transition from soft to
intermediate to hard state used in this analysis. {\scshape right} The
two best sampled outbursts studied by {\it RXTE}.  The x-axes have
been adjusted to show the similarity in the behaviour in the soft
states between the two outbursts.}
\label{fig:model}
\end{figure}

\noindent In Fig. \ref{fig:model} we show the standard Hardness Intensity
Diagram (HID) for all the observations of {\it GX~339-4}.  The
colour-scale and symbol type show the best fitting model for that
observation.  We also show the X-ray colours where the source
transitions between the hard, intermediate\footnote{We do not
  distinguish between the hard-intermediate and soft-intermediate
  states here.} and soft states.

The limitations of using {\it RXTE} data are apparent in the HID as
only when the disc is very bright is it easily fit in the spectrum.
The broken powerlaw points may well have a disc component, but the
curvature of the disc compared to the absorbed powerlaw is not
sufficient to cause the disc to be taken as the best fitting model in
the intermediate states.  The similarity in the transition fluxes
between some of the outbursts is clearly visible, especially on the
return to the hard state.

We also show in Fig. \ref{fig:model} the similarity between two of the
outbursts of {\it GX~339-4}.  Both the flux of the binary, shown as the
colour scale, and the X-ray colour through the outburst are very
similar, though the duration of the outbursts are in a ratio of 4:3,
and they are 2 years apart.  The hardening of the spectrum in the
middle of the outburst, where at the same time the flux increases, is
seen in all three well sampled outbursts.

\subsection{Discs and Lines}

\noindent As we fit disc and line components to all spectra, we can study the
evolution of these features as the outbursts progress.

\begin{figure}
\includegraphics[width=7.5cm]{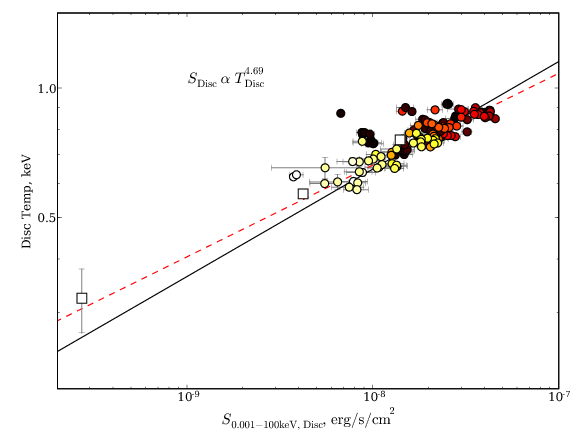}
\includegraphics[width=7.5cm]{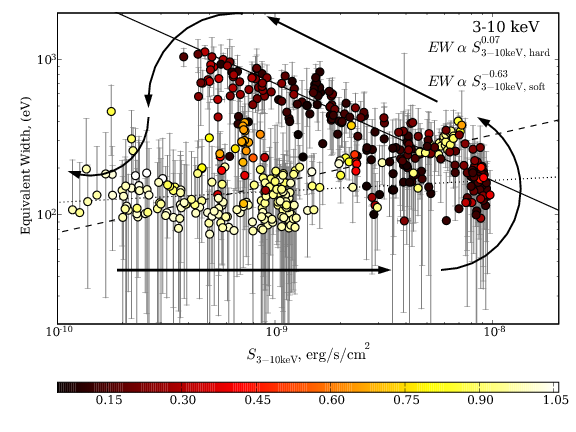}
\caption{{\scshape left} The change in disc flux as the disc
  temperature falls during the outburst.  The colour scale shows the
  date since the appearance of a disc component in the spectrum (dark
  is early, light is late).  The red dashed line is the best fit to
  the data points, and the black solid line is the theoretical
  prediction.  The square points are taken from \cite{Miller04a},
  \cite{Miller04b} and \cite{Miller06}. {\scshape right}  The change in
  equivalent width of a $6.4$~keV line component with flux.  The
  colour scale shows the X-ray colour of the observation.  The arrows
  show the motion
  through the diagram as the outburst progresses.}
\label{fig:discs}
\end{figure}

It is expected that discs follow a Stefan-Boltzmann like law where the
flux goes as
\[
S \propto T^4,
\]
as long as the geometry of the disc does not vary during the
outburst.  We show in Fig. \ref{fig:discs} the change in disc flux as the
temperature of the disc decays during the soft state.  Although the
best fit line does not quite match the theoretically expected
relation, they are qualitatively consistent.  The small range in
temperature and flux combined with the scatter in the relation are the
likely cause of the mis-match. 

In many of the observations we are able to fit a significant line and
extract reliable parameters for it.  The variation of the line
equivalent width (EW) with the flux is also shown in Fig. \ref{fig:discs}.
The colour scale shows the X-ray colour of the observation.  There
appears to be little evolution of the EW with flux in the hard state.
However, there is a clear anti-correlation in the soft state.  This is
very similar to the X-ray Baldwin effect observed in AGN.  However, in
{\it GX~339-4} the variation is observed within observations of one
object rather than from a study of many sources.

The AGN X-ray Baldwin effect has been mainly seen in studies of radio
quiet AGN \cite{Page, Iwasawa, Bianchi}.  In the standard model for X-ray binary outbursts derived
primarily from X-ray and Radio observations \cite{Fender04}, bright
radio emission is not expected or observed in the soft state.
Therefore the analogy between X-ray binaries and AGN is supported by
this Baldwin effect in {\it GX~339-4}.

\subsection{Links to AGN}

\noindent A number of studies have to date linked properties of X-ray binaries
and AGN together, where observed and derived parameters are scaled by,
for example, mass or accretion rate.  When trying to construct a more
general version of the HID to use for AGN, \cite{Koerding06} developed
the Disc Fraction Luminosity Diagram.  The luminosity of an X-ray
binary or AGN can easily be calculated.  The X-ray colour varies a
large amount during an X-ray binary's outburst, showing the relative
levels of the disc and powerlaw emission.  However in AGN, it
tends to give indications on to the presence and characteristics of a
warm absorber or other absorption features in the spectrum.  The
reason being that the AGN
disc's emission peaks in the UV, whereas in X-ray binaries the discs
are hotter and emit mainly in the soft X-ray band.  

Therefore generalise the X-ray colour axis to enable comparisons to
AGN to be performed, the Disc Fraction was calculated.  This
determines the amount by which the disc dominates the emission from
the source.  The X-ray colour does this very well for AGN, but it is
difficult to compare between sources, as the absorption cannot be
taken into account.  \cite{Koerding06} defined the disc fraction as
\[
{\rm Disc\, Fraction}=\frac{L_{0.1-100{\rm \ keV,\ PL}}}{L_{0.001-100{\rm
      \ keV,\ Disc}}+L_{0.1-100{\rm \ kev,\ PL}}}.
\]
In reality this is the powerlaw fraction, however rather than use the
true disc fraction (1-powerlaw fraction), we use this as the scales
change in the same way as in the HID.  This makes comparing the HID and
DFLD easier and re-training of the brain not a problem.

\begin{figure}
\includegraphics[width=7.5cm]{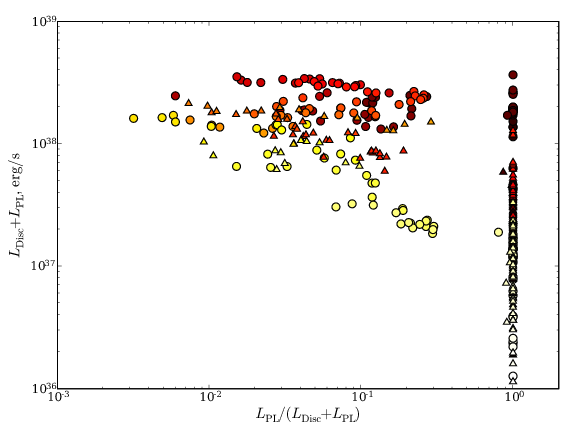}
\includegraphics[width=7.5cm]{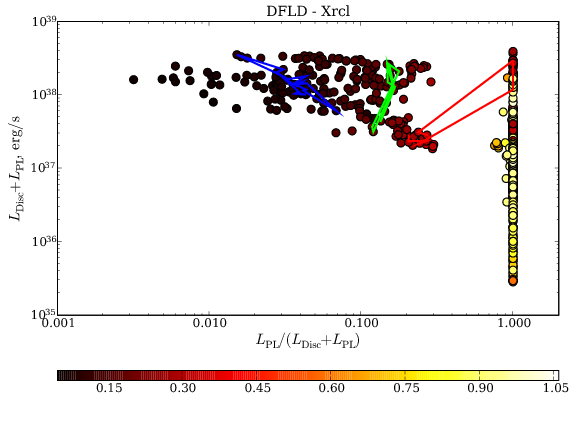}
\caption{{\scshape left} The DFLD for the two best covered
  outbursts. {\scshape right} The DFLD with the X-ray colour shown as
  the colour scale, and lines joining points with X-ray colours of
  {\scshape red} 0.3,  {\scshape green} 0.2,  {\scshape blue} 0.1.
\label{fig:dfld}}
\end{figure}

The disc and powerlaw luminosities are trivially extracted from the
spectral fitting routine and the DFLD is shown in
Fig. \ref{fig:dfld}.  We also show the mapping of the X-ray colour
onto the DFLD.  It is a non-linear mapping, expanding out the high
luminosity soft state and compressing the low luminosity regions.  We
also note that because of the limitations of the {\it RXTE} data,
non-dominant discs in the intermediate states are not well detected in
{\it GX~339-4}.

\section{Population Study}

\noindent As outlined in Section \ref{sec:intro} we have attempted to study as
many X-ray binaries whose outbursts were covered by {\it RXTE}.  Of
the 19 X-ray binaries analysed to date, we have observations of around
20 outbursts.  Not all binaries have outbursts which were observed by
{\it RXTE} and some have more than one.  For the remainder of this
analysis we use the Disc Fraction to indicate the states of the black
holes rather than the X-ray colour for the reasons mentioned above,
and also because the effects of galactic absorption can be accounted
for.  Previous work on comparing outbursts of X-ray binaries has been
done by \cite{Homan05,Remillard06}.

\subsection{DFLD}

\begin{figure}
\includegraphics[width=0.98\textwidth]{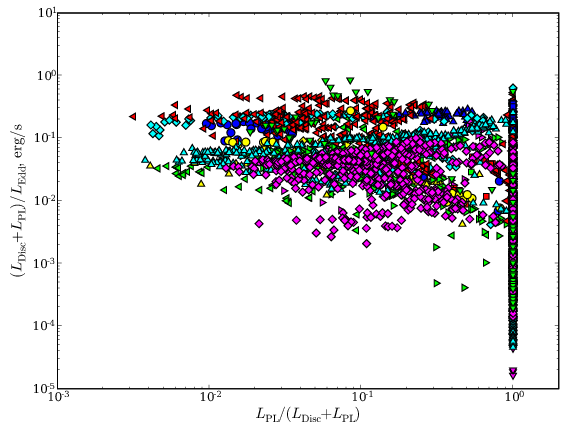}
\caption{The DFLD for all the X-ray binaries studied, each with their
  own symbol.}
\label{fig:dfld_all}
\end{figure}

\noindent In Fig. \ref{fig:dfld_all} we show the DFLD for all of the binaries
studied and for which reliable disc fractions could be calculated.
The general shape is as expected from the preliminary study of {\it
  GX~339-4}, however the gap seen between the powerlaw and disc
dominated states has been filled.  The shape of the diagram is vary
similar to that in \cite{Koerding06} (see their Fig. 10).

\begin{figure}
\includegraphics[width=7.5cm]{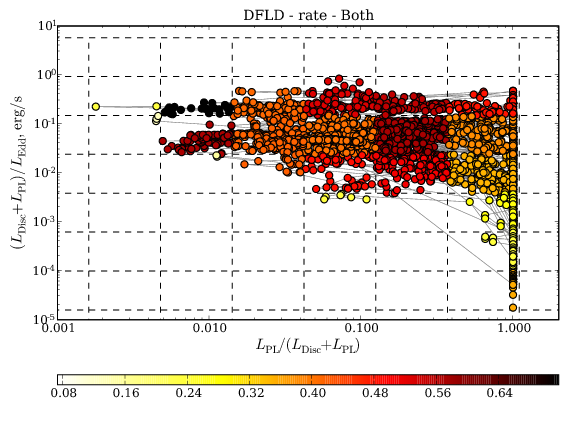}
\includegraphics[width=7.5cm]{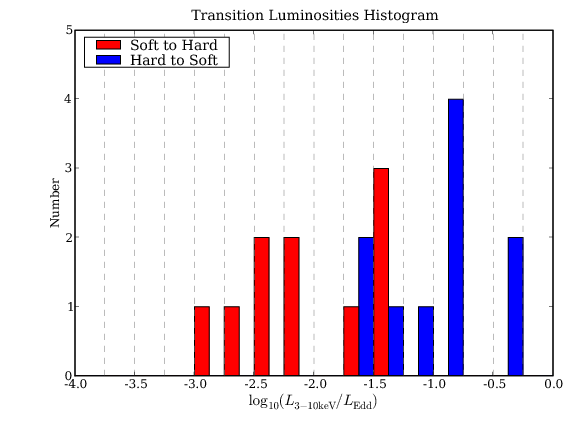}
\caption{{\scshape left} The rates of motion across the
  DFLD. {\scshape right} Histogram of the transition luminosities from
  and to the powerlaw dominated state.}
\label{fig:rates}
\end{figure}

We calculate the rate of motion through the DFLD and present in
Fig. \ref{fig:rates} the results, when averaged within a given region
of the DFLD.  There is surprisingly little variation in the rate
across the diagram, when taking into account that those regions with
the fewest points tend to have extreme values.  When splitting the
rates into luminosity or disc fraction changes, the reason is clear.
The movements in the two parameters nicely mirror each other.  When
there are large changes in the luminosity, the disc fraction is
relatively constant (powerlaw dominated state) and vice versa in the
disc dominated state.  Further investigation is required to see if
there is variation between objects or whether this quasi constant rate
of motion through the diagram results from looking at the average for
all objects.

Having analysed the spectra for all the outbursts we have data for, we
calculated the luminosities for the departure from and the return to
the powerlaw dominated state.  As can be seen in
Fig. \ref{fig:outbursts}, in some cases we only have observations
covering the last half of the outburst.  Although in these cases we
can easily determine the luminosity of the return to the powerlaw
dominated state, we do not include these in the histogram of
transitions shown in Fig. \ref{fig:rates}.  We only show the
luminosities for those outbursts where both the ingress to and egress
from the soft state are determined.  

\cite{Maccarone03} showed using the data at the time that the
transitions from the soft to the hard state occurred at similar
fractions of the Eddington rate (1-4\%).  They did restrict themselves
to objects where there were good mass, distance and flux estimates for
the transition times.  At this stage in our analysis we have not
restricted ourselves to objects which have these parameters well
defined.  However, there does appear to be some form of spread in both
the outbound and return transitions, rather than a narrow peak.

\subsection{Outbursts}

\begin{figure}
\includegraphics[width=7.5cm]{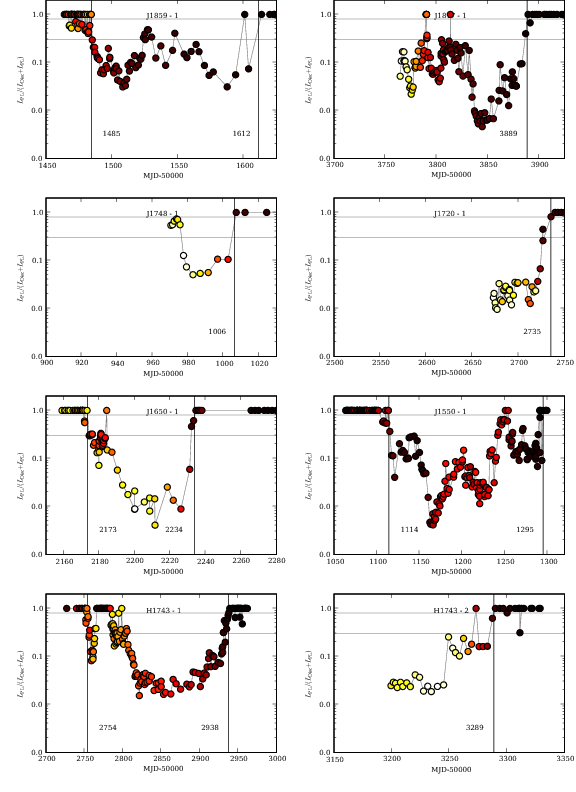}
\includegraphics[width=7.5cm]{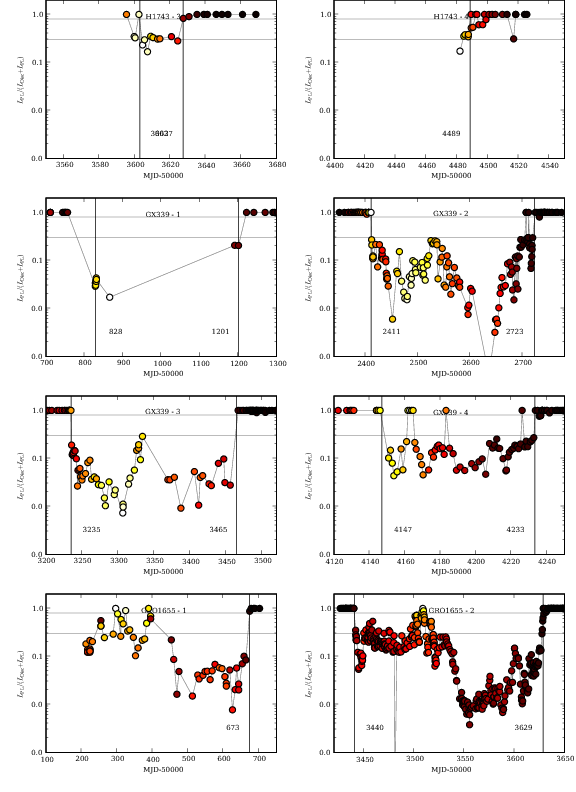}
\caption{A selection of the Disc Fraction - time curves for the
  outbursts in the sample.  The colour scale is for the luminosity of
  the source during the outburst}
\label{fig:outbursts}
\end{figure}

\noindent As the two outbursts presented in Fig. \ref{fig:model} were so similar, we
investigated whether similarities existed between outbursts from other
X-ray binaries.  A selection of the outbursts analysed are shown in
Fig. \ref{fig:outbursts}.  The horizontal lines show the disc
fractions used to delineate the powerlaw and disc dominated states.
The vertical lines show when a transition has been deemed to occur.  In
a number of the well sampled outbursts, the hardening of the spectrum
in the middle of the disc-dominated state and the associated
brightening of the source is seen.   Even in some of the outbursts
where only the tail end is observed, there are hints that this feature
is also present.

%\subsection{Discs and Lines}

%\begin{figure}
%\centering
%\includegraphics[width=0.6\textwidth]{Lum_EW_DF_bw.png}
%\caption{A selection of the Disc Fraction - time curves for the
%  outbursts in the sample.  The colour scale is for the luminosity of
%  the source during the outburst}
%\label{fig:lines}
%\end{figure}

\section{Summary}

\noindent We have analysed 14Ms of {\it RXTE} data on 19 black-hole X-ray
binaries, with the aim of studying their outbursts as well as the disc
and line properties.  We have presented preliminary results from this
study, including a new diagnostic diagram derived from the
luminosities of the different spectral components.  Using this disc
fraction we are able to take into account the differing galactic
absorptions for different objects.  We show that the rate of motion
through the diagram is fairly uniform.  There are also clear
similarities between outbursts of different objects when the variation
in disc fraction is plotted against the date of the observation.
These disc-fraction curves also let the transitions from hard to soft
and vice versa to be calculated and we show the distribution of
transition luminosities.


\begin{thebibliography}{99}
\bibitem{Bianchi} S.~Bianchi, M.~Guainazzi, G.~Matt, N.~Fonseca
  Bonilla, 2007, \emph{On the Iwasawa-Taniguchi effect of radio-quiet
    AGN}, \emph{A\&A}, 467, 19
\bibitem{Dunn08} R.~J.~H.~Dunn, R.~P.~Fender, E.~K\"ording, C.~Cabanac \& T.~Belloni, 2008,
  \emph{Studying the X-ray hysteresis in GX 339-4: the disc and iron line over one decade}, \emph{MNRAS}, 387, 545
\bibitem{Fender04} R.~P.~Fender, T.~M.~Belloni, E.~Gallo, 2004,
  \emph{MNRAS}, \emph{Towards a unified model for black hole X-ray
    binary jets}, 355, 1105
\bibitem{Homan05} J.~Homan, T.~M.~Belloni, 2004,
  \emph{Ap\&SS}, \emph{The evolution of black hole states}, 300, 107
\bibitem{Iwasawa} K.~Iwasawa, Y.~ Taniguchi, 1993 ,\emph{The X-ray
  Baldwin Effect}, \emph{ApJ}, 413, 151
\bibitem{Koerding06} E.~K\"ording, S.~Jester \& R.~Fender, 2006,
  \emph{Accretion states and radio loudness in active galactic nuclei:
    analogies with X-ray binaries}, \emph{MNRAS}, 372, 1366
\bibitem{Maccarone03} T.~J.~Maccarone, 2003, \emph{Do X-ray binary
  spectral state transition luminosities vary?}, \emph{A\&A}, 409, 697
\bibitem{Miller04a} J.~M.~Miller \emph{et al.}, 2004, \emph{Evidence of Black Hole Spin in GX 339-4: XMM-Newton/EPIC-pn and RXTE Spectroscopy of the Very High State}, \emph{ApJ}, 606, L131
\bibitem{Miller04b} J.~M.~Miller \emph{et al.}, 2004, \emph{Chandra/High Energy Transmission Grating Spectrometer Spectroscopy of the Galactic Black Hole GX 339-4: A Relativistic Iron Emission Line and Evidence for a Seyfert-like Warm Absorber} \emph{ApJ}, 601, 450
\bibitem{Miller06} J.~M.~Miller \emph{et al.}, 2006, \emph{A Long, Hard Look at the Low/Hard State in Accreting Black Holes}, \emph{ApJ}, 653, 525
\bibitem{Page} K.~L.~Page \emph{et al.}, 2004, \emph{An X-ray Baldwin
  effect for the narrow Fe Ka lines observed in active galactic
  nuclei}, \emph{MNRAS},  347, 316
\bibitem{Remillard06} R.~A.~Remillard, J.~E.~McClintock, 2006,
  \emph{X-ray Properties of Black-Hole Binaries}, \emph{ARA\&A},  44, 49

\end{thebibliography}
\end{document}